# Computational Model to Quantify Object Innovativeness

**Vladimir K Ivanov**

Centre of New Information Technologies, Tver State Technical University, Afanasy Nikitin emb. 22, Tver, Russia

**Abstract.** The article considers the quantitative assessment approach to the innovativeness of different objects. The proposed assessment model is based on the object data retrieval from various databases including the Internet. We present an object linguistic model, the processing technique for the measurement results including the results retrieved from the different search engines, and the evaluating technique of the source credibility. Empirical research of the computational model adequacy includes the acquisition and preprocessing of patent data from different databases and the computation of invention innovativeness values: their novelty and relevance. The experiment results, namely the comparative assessments of innovativeness values and major trends, show the models developed are sufficiently adequate and can be used in further research.

## 1. Introduction

The analysis of the literature dedicated to different innovation issues (see, e.g., [1], [2], [3]) shows that the notion 'innovation' always includes such connotations as 'new', 'efficiency-increasing', 'profit-making'. The object innovation is considered within its system relations, the degree of the object influence on a subject and external objects.

The article examines the quantitative assessment approach to the innovation of various objects such as engineering solutions and system, product, technology components.

The proposed assessment model is based on the object data retrieval from various databases including the Internet. The paper presents key approach concepts, general indicators of innovativeness assessment, an object linguistic model, a technique to process the measurement results including the results retrieved from different search engines, and a technique to evaluate the source credibility. We introduce the specific application of the theory of evidence for combining the measurement results of different indicator sources including the Internet. The validity of the proposed models is verified experimentally. The experiments results shown in the article make it clear that our approach is rational and can be used to analyze products and technologies, to make business solutions, to evaluate projects.

It should be mentioned that some object properties also defining innovation in general are beyond the scope of the present article, with the availability of a patent (or patentability), cost-effectiveness, and judicial admissibility being an example.

## 2. Key Concepts and Terms

Below you can find key terms and notions as they are understood and identified in the article.

*Innovation* is an object having some properties determining its technological novelty, relevance, and implementability. Innovation may be an invention, an engineering system component, a technology, a method, etc.

*Technological novelty* means significant improvements, a new way of using or granting a product (a system component) or a technology. Novelty subjects are potential users or producers themselves.

*Relevance* is a potential producer's awareness of the object necessity as a demand.

*Implementability* of a product determines the technological validity, physical feasibility and capacity of a product to be integrated in the system in order to have desired effects.

*Target object* is an object whose information is searched for in a database with a search engine. In our context a target object is a product (technology) with a possible innovation potential.

*Search pattern* is a target object description (a linguistic model) consisting of a marker and key terms. A search pattern is used to generate queries.

*Key term* is a word or a phrase determining a target object in one of the following ways:
- an application mechanism (operation), an object structure;
- an object application (operation) result or conditions;
- important object characteristics (properties, material, composition).

*Marker is* a key term determining an object application (action) field.

*Query* is a set of keywords and a marker used by a search engine to search object information in a database.

*Search engine* is a mechanism of searching information in a database.

## 3. Simulating Object Innovativeness Indicators

*3.1. Using an Object Linguistic Model*

An object linguistic model consists of $N$ key terms. Suppose there is a universal set of term numbers $\Omega=|N|$. For example, if $N=4$, then $\Omega=\{1,2,3,4\}$. The marker has number $0\notin\Omega$. Queries are constructed with the terms $A_{qk}\subseteq\Omega$, $k=[1,N]$.

For example, $A_{q1}=\{1\}$, $A_{q2}=\{1,2\}$, etc. A marker is added to each query. The total number of queries $S$ is determined as the combination of elements $a$ to $b$ without repetition which is estimated as follows:

$$S = 1 + \sum_{c=1}^{N-1} N!/c!(N-c) \quad (1)$$

where $c$ is the number of terms in a query, and the unity in the expression is identified with a query where $c=N$. For example, if $N=4$, then $S=15$.

Queries $A_{qk}$ are executed in a search engine and serve as observable subsets $\Omega$ (focal elements). The query execution results in determining a set of database entries (documents for document databases) retrieved with power R.

*3.2. Assessing Innovativeness Indicators*

The assessment model of innovativeness indicators is based on the object data retrieval from various databases including the Internet and given below.

An object novelty $Nv$ is determined as follows:

$$Nv = 1 - \frac{1}{S}\sum_{k=1}^{S} R_k^{01}, \quad R_k^{01} = 1 - \exp(1 - \frac{R_k}{R_0^{01}}) \quad (2)$$

where $R_k$ is a number of documents resulted from the execution of $k$-th query in a database containing the information about the data domain under consideration; $R_k^{01}$ is the value of $R_k$ standardized for $[0;1]$; $R_0^{01}$ is the number of documents standardized for $[0;1]$ and resulted from the execution of a query containing one term-marker.

The number of new object search results being applicable to a search pattern are supposed to be less than that of old and well-known ones.

Object relevance $Rl$ is estimated as follows:

$$Rl = \frac{1}{S}\sum_{k=1}^{S} F_k^{01}, \quad F_k^{01} = 1 - \exp(1 - \frac{F_k}{F_0^{01}}) \quad (3)$$

where $F_k$ is a quantity showing the users' interest to the target object. The quantity can be:
- the frequency of users' queries similar to *k*-th query;
- the number of users' citations of the target object description.

It should be noted that different interpretations of $F_k$ are possible.

$F_k^{01}$ is the value of $F_k$ standardized for [0;1]; $F_0^{01}$ is the value of $Fk^{01}$ for the query containing one term-marker.

The number of relevant object search results being applicable to a search pattern and a linguistic model are supposed to be larger than that of the objects with demand decline. Table 1 presents estimated and measured indicators of query executions $A_{qk}$.

**Table 1.** Estimated and measured innovativeness indicators.

| Estimated indicators | | Measured indicators | |
|---|---|---|---|
| Notation | Description | Notation | Description |
| *p(Nv)* | The probability of an object having technological novelty | *R* | The number of documents retrieved with the search engine |
| *p(Rl)* | The probability of an object having relevance (in demand) | *F* | The frequency of search engine users' queries or The number of search engine users' citations of the retrieved target object description |

## 4. Data Processing Techniques

*4.1. Data Processing Technique for Measurement Results*

Let us consider the order of computation for innovation indicator *p(Nv)* as an example.

The number of the retrieved documents $R_k$ is computed for each query and specified in accordance with (2). For all $R_k$ the number of group intervals is determined as $I = S^{1/2}$. Then for the example in question when $S=15$, $I = 15^{1/2} = 3.873 \approx 4$ in case of the equal intervals $A_1=[0.0;0.25]$, $A_2=[0.26;0.50]$, $A_3=[0,51;0,75]$, $A_4=[0.76;1.0]$. In terms of measuring *Nv* the mentioned intervals correspond to the nominal scale "It is novel", "It is evidently novel", "It is evidently not novel", "It is not novel".

Suppose that query executions using Search Engine 1 (SE1) give the following results for indicator $R_k$: three queries have the results from the interval $A_1$ ($q_1=3$), four queries have the results from the interval $A_2$ ($q_2=4$), seven queries have the results from the interval $A_3$ ($q_3=7$), one query has the result from the interval $A_4$ ($q_4=1$).

According to the theory of evidence [4] base probability $m(A_k)$ or frequency function is determined as:

$$m: P(\Omega) \to [0;1](2^\Omega \to [0;1]), m(\varnothing) = 0, \sum_{A \in P(\Omega)} m(A) = 1 \qquad (5)$$

and can be estimated as follows:

$$m(A_k) = q_k / S \qquad (6)$$

where $q_k$ is a number of observed subsets (queries), $\sum q_k = S$.

Table 2 presents the results of the calculations of $m(A_k)$.

Further a belief function $Bel(A) = \sum_{A_k : A_k \subseteq A} m(A_k)$ and a plausibility function $Pl(A) = \sum_{A_k : A_k \cap A} m(A_k)$ have to be estimated. The estimation results for the intervals $A_1$ and $A_2$ are the following: $Bel(A)=m(A_1)+m(A_2)=0.47$, $Pl(A)=0$. Thus the observed object is new with probability $p(Nv)=0.47$.

**Table 2.** The results of the base probabilities calculations.

| Interval $A_k$ | Number of queries $q_k$ | Base probability $m(A_k)$ |
|---|---|---|
| 1 | 3 | 0.20 |
| 2 | 4 | 0.27 |
| 3 | 7 | 0.46 |
| 4 | 1 | 0.07 |

Note that the sequence intervals ($A_k$) need not be equal. To identify the interval lengths precisely we should use the frequency distribution of values $R_k$.

*4.2. Processing Technique for Measurement Results Retrieved from Different Search Engines*

Suppose that Search Engine 2 (SE2) gives the measurement results for indicator $Rk$ when executing the same queries. Table 3 shows the results of the calculations of $m(A_k)$ for SE1 and SE2 together.

**Table 3.** The results of the base probabilities calculations for SE1 and SE2.

| Interval $A_k$ | SE1 | | SE2 | |
|---|---|---|---|---|
| | Number of queries $q_k$ | Base probability $m(A_k)$ | Number of queries $q_k$ | Base probability $m(A_k)$ |
| 1 | 3 | 0.20 | 6 | 0,40 |
| 2 | 4 | 0.27 | 6 | 0,40 |
| 3 | 7 | 0.46 | 1 | 0,07 |
| 4 | 1 | 0.07 | 2 | 0,13 |

The estimation results for the intervals $A_1$ and $A_2$ are the following: $Bel(A)=0.80$, $Pl(A)=0$. That is, $p(Nv)=0.80$ which differs from the results retrieved with SE1.

The combined base probability $m_{12}$ is calculated by formula:

$$m_{12}(A) = \frac{1}{1-K} \sum_{A_i^{(1)} \cap A_j^{(2)} = A} m_1(A_i^{(1)}) m_2(A_j^{(2)}) \qquad (7)$$

where the conflict factor $K$ is calculated as follows:

$$K = \sum_{A_i^{(1)} \cap A_j^{(2)} = \emptyset} m_1(A_i^{(1)}) m_2(A_j^{(2)}) \qquad (8)$$

Table 4 presents data intersections example for SE1 and SE2 mentioned above.

**Table 1.** Data intersections example.

| | | $A_j^{(2)}$ | | | |
|---|---|---|---|---|---|
| | | {8,9,10,11,12,13} | {2,4,5,6,14,15} | {1} | {3,7} |
| $A_i^{(1)}$ | {1,2,3} | ∅ | {2} | {1} | {3} |
| | {4,5,7,8} | {8} | {4,5} | ∅ | {7} |
| | {6,9,10,11,12,13,14} | {9,10,11,12,13} | {6,14} | ∅ | ∅ |
| | {15} | ∅ | {15} | ∅ | ∅ |

In accordance with (8) $K=0.23$, and according to (6) $m_{12}(A_1)=0.15$, $m_{12}(A_2)=0.33$, $m_{12}(A_3)=0.48$, $m_{12}(A_4)=0.04$. The estimation results for the intervals $A_1$ and $A_2$ are the following: $Bel_{12}(A)=0.48$, $Pl_{12}(A)=0$, $p(Nv)=0.48$.

*4.3. Evaluating Technique of Source Credibility*

The source credibility can be considered with the introduction of discount factor $\alpha$ for base probability $m(A)$ [4]. Suppose the expert has assigned $\alpha_1=0$ for SE1 and $\alpha_2=0.2$ for SE2. Hence, the results of the first engine are more credible that those of the second one. Discounted base probabilities are estimated as follows:

$$m^\alpha(A) = (1-\alpha)m(A) \qquad (9)$$

With Dempster's rule and expressions (7) and (8) for $m_1^\alpha(A_i^{(1)})$ и $m_2^\alpha(A_j^{(2)})$ we have $K^\alpha=0.19$, $m_{12}^\alpha(A_1)=0.12$, $m_{12}^\alpha(A_2)=0.25$, $m_{12}^\alpha(A_3)=0.36$, $m_{12}^\alpha(A_4)=0.03$. The estimation results for the intervals $A_1$ and $A_2$ are the following: $Bel_{12}(A)=0.37$, $Pl_{12}(A)=0$, $p(Nv)=0.37$.

*4.4. Data Processing Technique for Linguistic Model*
The task of making an object linguistic model consists in defining N key terms for the target object. It can be done by an expert either manually or with some means of automatic key word extraction. There are many methods of key word, phrase and notion extraction from the texts of different types (see, e.g. review [5]). Today the problem goes on to be relevant. The number of publications connected with 'key word extraction' in the database of scientific articles ACM DigitalLibrary (https://dl.acm.org) has been steadily rising in recent years: 2015 year – 2819 articles, 2016 – 3177, 2017 – 3838.

Our task has several peculiarities that are worth to be mentioned:
- the number of terms in an object linguistic model should be small in order to limit value $S$ being calculated according to (1);
- source object descriptions can be of different origins, structures, and volumes, so the method of key term extraction should be invariant to source texts;
- for the same reason, the method of key term extraction should be able to adapt to source texts, i.e. object descriptions.

Based on the peculiarities mentioned we suggest and validate an evolutionary approach to the solution of an object linguistic model problem [6]. The general concepts of suggested approach are described below. It is the technology of generating search queries, filtering and ranking search results. The main idea is to organize with a special genetic algorithm an evolutionary process generating a stable and effective system query population for getting highly relevant results. In the course of the process coded queries are sequentially exposed to genetic changes and made in a search engine. Then the semantic relevance of intermediate search results is evaluated, fitness function values are computed, and the most appropriate queries are selected.

The search pattern of document $K$ is a set of key words from text documents being reference ones for the subject search area.

Each search query is coded with vector $\overline{q}=(c_1,c_2,...,c_n,...,c_m)$, where $c_n=(k_n, S_n)$, $k_n \in K$ is a term, $S_n$ is a set of synonyms for term $k_n$. The result of a search query is a set of documents $R, |R|=D$. The set $R$ is grouped after performing $\overline{q}$ in a search engine (Bing, Google, Yandex, SQL database, XML-data, etc.).

The initial population from $N$ search queries is presented as a set of $Q_0$, where $|Q_0|=N, N<|K|/2, \overline{q} \in Q_0$. The crossover (one-point or two-point) is carried out by exchanging the terms between components of vectors $\overline{q}$, genotype outbreeding being used for query reproduction. The most adequate mutation operation is the probabilistic change of query term $k_n$ chosen randomly by synonym $k_n' \in S_n$. To generate a new query population an elite selection is used. Generally, the condition of terminating the algorithm is considered to be population stability.

In our case, the highly relevant results are those ones fitting with the larger values of a fitness function. So, the value of the fitness function $R_i$ for the $i$-th result of the $j$-th query (i. e. for result $r_i$) is determined from the following:

$$R_i = w_g * g + w_p * p + w_s * s \tag{10}$$

where value $g$ takes account of rank for $r_i$ set by a search engine; value $p$ takes account of genericity $r_i$ that is frequency of occurrence $r_i$ in the list of results of other queries; value $s$ determines semantic similarity $r_i$ and search pattern $K$ (we use a cosine semantic similarity measure of document vectors as it is common in a vector space model [7]); $w_g$, $w_p$ and $w_s$ –weighting factors for $g$, $p$ and $s$ correspondingly. Note, the values $g$, $p$ and $s$ are standardized for [0;1].

As a result of an evolutionary process we get a set of effective queries $Q_N = \{q_n = \{c_{nk}\}\}$ giving the most relevant results. A multiset $C_N = \{c_{nk} \in C_N \mid c_{nk} \in K, c_{nk} \in q_n, q_n \in Q_N\}$ is interpreted as a linguistic model of a target object. The weight and significance of $w_n$ term is determined by the multiplicity of a term, i.e. the number of queries with term $c_{nk}$.

## 5. Results of experiments

*5.1. Purpose of experiments*

This section presents the results of the made experiments. The purpose of the experiments is as follows:
- To calculate object innovativeness indicators following the adopted computational model (see Section 3).
- To compare the computed results of innovativeness indicators with their expert reviews.
- To compare the computed results of innovativeness indicators obtained after processing data from different search engines.
- To assess the dynamic changes of object innovativeness indicators in time.
- To make sure that the obtained measurement data of innovativeness indicators are suitable for further processing with data processing techniques (see Section 4).

*5.2. Source data*

The ten best Russian projects of 2017 chosen by the experts of Rospatent (Federal Service for Intellectual Property, https://rupto.ru) have become assessment objects. The information about the projects is presented at https://ru.rt.com/9zqx. Table 5 shows the list of the patents (further TOP-10) with their linguistic model components. The patent descriptions are taken from http://new.fips.ru/en. Expert reviews are used to determine markers and key terms.

**Table 5.** Patents TOP-10 and their linguistic models.

| Seq. # | Patent # | Marker | Key terms |
|---|---|---|---|
| 1 | RU 2620154 C1 | eye | "optic nerve" "polymer base" treatment electrostimulation |
| 2 | RU 2623171 C1 | regeneration | "gene-activated material" "matrix-carrier" "nucleic acid" biotechnology |
| 3 | RU 2632806 C1 | heart | implantation minicontour "childhood" hematolysis |
| 4 | RU 2624867 C1 | parodontium | "dental plaque" ultrasound "antiseptic treatment" "dental pencil" |
| 5 | RU 2639015 C1 | blockchain | "product quality" "information flow" "information protection" "distributed database" |
| 6 | RU 2614230 C1 | 3D printing | "precision alloy" niobium microhardness "wear resistance" |
| 7 | RU 177015 U1 | space | refuse clearing orbit spacecraft |
| 8 | RU 2641323 C1 | rocket | "low thrust" "liquid engine" reliability thriftiness |
| 9 | RU 2642701 C2 | wood | "heat treatment" "gas fuel" "water steam" circulation |
| 10 | RU 176787 U1 | slope | level measurement "LED indicator" "three-axis accelerometer" |

Source information was obtained and innovativeness indicators were computed with the following search engines and appropriate databases:

- http://new.fips.ru
- https://elibrary.ru
- https://rosrid.ru
- https://yandex.ru
- https://wordstat.yandex.ru
- https://google.com
- https://adwords.google.com
- https://patents.google.com
- https://scholar.google.ru

*5.3. Results of experiments*

Figure 1 shows the example of TOP-10 patent novelty computations made in accordance with (2) and with two databases (for patents see table 5).

Table 6 shows mean novelty values for the TOP-10 patents and the ten patents registered in 2017 and randomly taken out from http://new.fips.ru database.

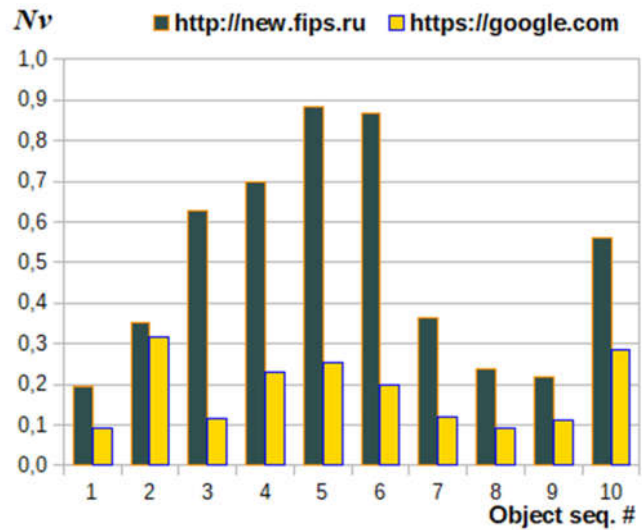

**Figure 1.** An example of object novelty computations.

**Table 6.** Mean Value $\overline{Nv}$ and Standard Deviation $\sigma_{Nv}$ of novelty for patent groups.

|  | http://new.fips.ru | | https://google.com | |
| --- | --- | --- | --- | --- |
|  | TOP-10 patents | 10 random patents | TOP-10 patents | 10 random patents |
| $\overline{Nv}$ | 0.50 | 0.37 | 0.18 | 0.23 |
| $\sigma_{Nv}$ | 0.26 | 0.12 | 0.08 | 0.15 |

Figure 2 shows the example of TOP-10 patent relevance computations made in accordance with (3) and with two other databases.

Table 7 shows mean relevance values for TOP-10 patents and ten patents registered in 2017 and randomly taken out from http://new.fips.ru database.

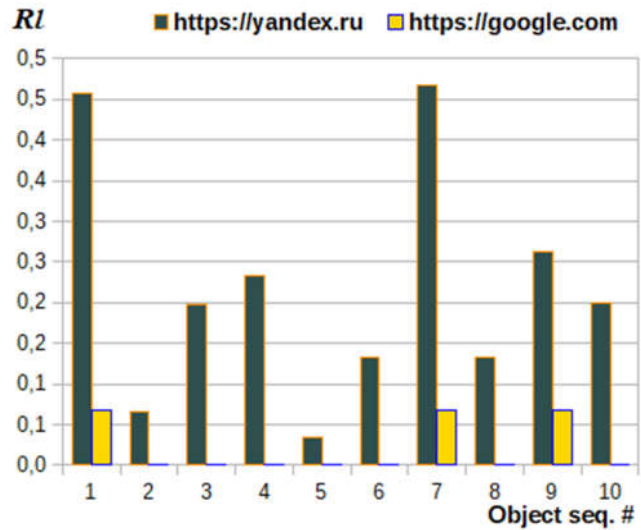

**Figure 2.** An example of object relevance computation results.

**Table 7.** Mean Value $\overline{Nv}$ and Standard Deviation $\sigma_{Nv}$ of relevance for patent groups.

|  | https://yandex.ru | | https://google.com | |
|---|---|---|---|---|
|  | TOP-10 patents | 10 random patents | TOP-10 patents | 10 random patents |
| $\overline{Rl}$ | 0.22 | 0.26 | 0.02 | 0.01 |
| $\sigma_{Rl}$ | 0.15 | 0,14 | 0.03 | 0.03 |

Figure 3 shows the novelty computation results of patent RU 2624867 C1 (see table 5) made in accordance with (2) by each year from 1998 to 2017 with two databases.

Figure 4 shows the relevance computation results of patent RU 2624867 C1 made in accordance with (3) by each year from 1998 to 2017 with the same databases.

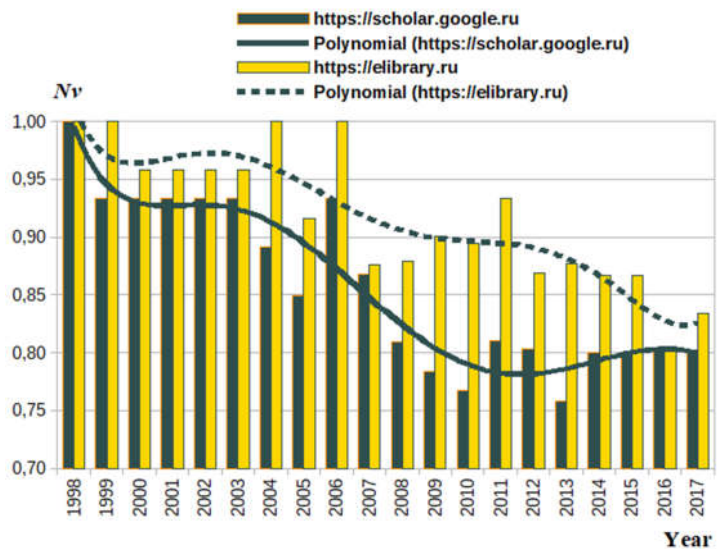

**Figure 3.** The novelty of the linguistic model of patent RU 2624867 C1 by years.

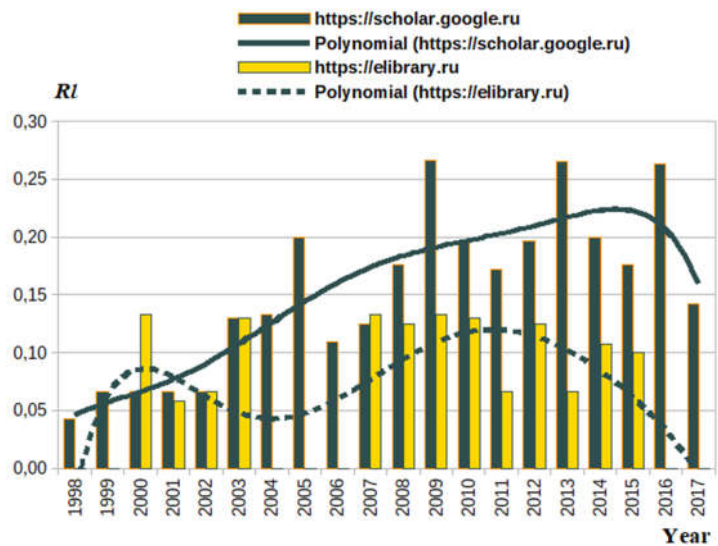

**Figure 4.** The relevance of the linguistic model of patent RU 2624867 C1 by years.

## 6. Discussion

The analysis of the experimental results has allowed us to draw the following findings:
1. When innovativeness indicators of specific objects are computed, absolute result values are expectedly different for two different databases. Our model is elaborated with the situation taken into account and suggests the theory of evidence to combine measurements results obtained from different sources with different data accuracy and reliability. See Subsections 4.2 and 4.3.
2. On the other hand, the trends identified show similar results of our computational model for different databases. See the diagrams of object novelty and relevance change in figure 1 and figure 2 respectively. It means that the model adequately assess relative changes of indicator values computed with different source data.
3. The innovativeness indicator values for different objects computed with some database differ from each other within the limits justified for normal value distribution (3-sigma rule). See mean values and standard deviations in tables 6 and table 7 respectively. It makes it possible to compare the assessment of indicator values reasonably.
4. Mean novelty values of TOP-10 patents are larger than mean novelty values of randomly selected objects (see table 6). Thus, the model under discussion proves the expert appraisals of innovativeness.
5. Similarly, Mean relevance values of TOP-10 patents are larger than mean relevance values of randomly selected objects (see table 7). It means that our model also proves the expert appraisals of innovativeness.
6. Some conclusions can be made with the analysis of patents novelty assessed for a certain period. The patent novelty was determined individually with the database subsets corresponding to patent registration years (a twenty-year period was used). Approximation of the obtained values $Nv$ verifies the hypothesis of the object novelty reduction in time (see figure 3).

Note that an innovativeness value is actually computed not for some specific object (e.g. patent RU 2624867 C1) but for some abstract object defined by a patent linguistic model. The abstract object has significant properties being common to the properties of the object in question. It is normal for our research since our objective here is to show the dynamics of novelty value change in time.

7. The relevance value of patent RU 2624867 C1 (or more exactly the object defined by a linguistic model of patent RU 2624867 C1) grows in time. The diagram in figure 4 shows the trend. It is also quite natural. During its existence the object gains more popularity among users so the potential interest in it grows.
8. Consider the time intervals in figure 3 and figure 4. We can see some definite cycle in innovativeness values. E. g., the evident peaks of the analyzed object relevance (the years of 2000 and 2007-2010) on database https://elibrary.ru as well as the peak of the analyzed object novelty (2000-2003) on database https://scholar.google.ru. Moreover, we see the correlation between the series of object novelty and relevance values. Meanwhile the computation results are ambiguous. The examples mentioned above show both strong correlation $corr(Nv, Rl) = 0.89$ (for database https://scholar.google.ru) and weak correlation $corr(Nv, Rl) = 0.28$ (for database https://elibrary.ru).

In any case, we identify the cyclicity that requires the check of the hypothesis of innovation cycles in this application area. It should be noted that the concepts describing the society development in general and economy in particular as a consequence of recurring cycles or long waves have long been recognized [8]. Our research only deals with the hypothesis generation which needs to be followed up.

## 7. Conclusions

The approach to determining object innovation indicators that has been described in the present article is one of the outcomes of the research Data Warehousing Based on Search Agent Intellectualization and Evolutionary Model of Target Information Selection. Object innovation indicators must serve as an important factor of selecting relevant information for target topic storage segments. We assume that the computational models developed are adequate and will be used further in the project. Carrying on this line of research we are planning to substantiate and formalize the generation procedure of an object linguistic model and include algorithms based on the examined techniques in a search agent composition.

**Acknowledgments**
This work was supported by Russian Foundation for Basic Research (RFBR), project No. 18-07-00358. The author is especially grateful to the team who has taken part in the project.